\documentclass[twocolumn,prl]{revtex4-1}
\usepackage[latin9]{inputenc}
\usepackage{float}
\usepackage{bm}
\usepackage{amsmath}
\usepackage{amssymb}
\usepackage{graphicx}

\makeatletter

\usepackage{bm}
\renewcommand{\textendash}{--}

\makeatother

\begin{document}

\title{Dirac points merging and wandering in a model Chern insulator}

\author{Miguel Gonçalves$^{1}$, Pedro Ribeiro$^{1,2}$, Eduardo V. Castro$^{1,2,3}$}

\affiliation{$^{1}$CeFEMA, Instituto Superior Técnico, Universidade de Lisboa,
Av. Rovisco Pais, 1049-001 Lisboa, Portugal}

\affiliation{$^{2}$Beijing Computational Science Research Center, Beijing 100084,
China}

\affiliation{$^{3}$Centro de F\'{\i}sica das Universidades do Minho e Porto,
Departamento de F\'{\i}sica e Astronomia, Faculdade de Ciências, Universidade
do Porto, 4169-007 Porto, Portugal}
\begin{abstract}
We present a model for a Chern insulator on the square lattice with
complex first and second neighbor hoppings and a sublattice potential
which displays an unexpectedly rich physics. Similarly to the celebrated
Haldane model, the proposed Chern insulator has two topologically
non-trivial phases with Chern numbers $\pm1$. As a distinctive feature
of the present model, phase transitions are associated to Dirac points
that can move, merge and split in momentum space, at odds with Haldane's
Chern insulator where Dirac points are bound to the corners of the
hexagonal Brillouin zone. Additionally, the obtained phase diagram
reveals a peculiar phase transition line between two distinct topological
phases, in contrast to the Haldane model where such transition is
reduced to a point with zero sublattice potential. The model is amenable
to be simulated in optical lattices, facilitating the study of phase
transitions between two distinct topological phases and the experimental
analysis of Dirac points merging and wandering.
\end{abstract}
\maketitle

\section{Introduction}

The study of topological phases in electronic systems, particularly
topological insulators, have become an area of vast interest in condensed
matter physics \citep{RevModPhys.82.3045,QZrmp11,bernevigBook}. Arguably,
the first experimenal realization of a topological electron system
is associated with the discovery of the quantum Hall effect by von
Klitzing \citep{Klitzing1980} almost fourty years ago. This phenomenon
takes place in two-dimensional systems at high magnetic fields, when
a quantized Hall conductivity and dissipationless surface conducting
states appear. High magnetic fields are not easy to realize, which
motivates the interest for an effect with the same transport properties
under a zero applied magnetic field \textendash{} the quantum anomalous
Hall effect \citep{NSO+10}.

Both quantum Hall and quantum anomalous Hall systems are two-dimensional
insulators with broken time-reversal symmetry. Their topological character
is associated with a topological invariant called first Chern number
($C$), which is equal to the Hall conductivity in units of $e^{2}/h$
\citep{Xiao2010}. Different Chern numbers describe different topological
phases of matter, while $C=0$ corresponds to the normal insulating
phase. In modern language this systems are known as Chern insulators
\citep{bernevigBook}.

\begin{figure}
\begin{centering}
\includegraphics[width=0.85\columnwidth]{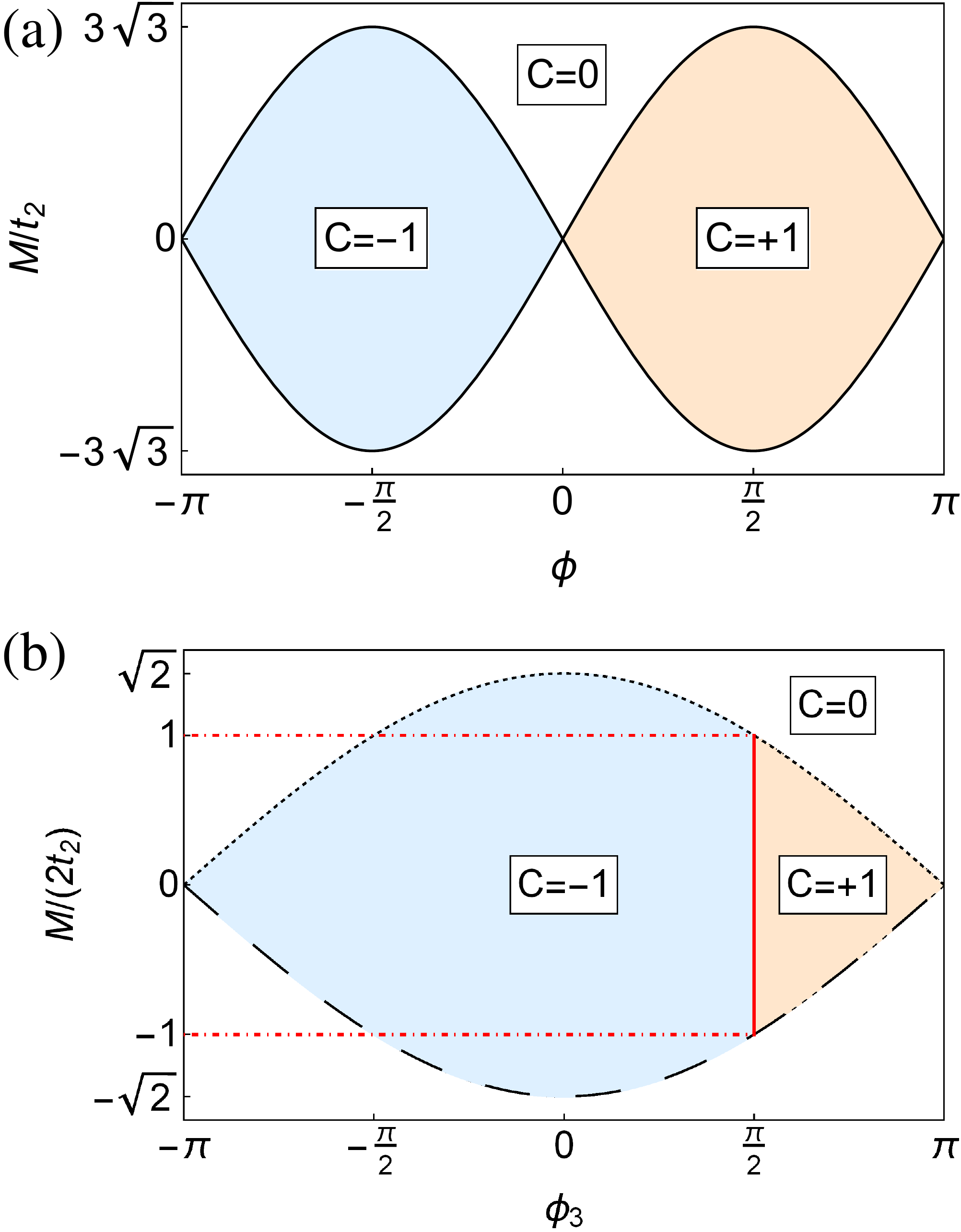}
\par\end{centering}
\caption{\label{fig:phase_diagram}(a) Phase diagram of the Haldane model.
(b) Phase diagram of the model proposed in Eq.~\ref{eq:Ham}.}
\end{figure}

The first theoretical realization of the quantum anomalous Hall effect
is due to Haldane~\citep{Haldane1988}. He proposed a Chern insulator
toy model considering a honeycomb lattice with two different types
of atoms subjected to a sublattice potential $\pm M$ and with second
nearest neighbor complex hoppings of the type $t_{2}e^{\pm i\phi}$.
These hoppings were responsible for breaking time reversal symmetry
and are the effect of a local magnetic flux density built in such
a way that its total flux in the unit cell, and therefore in the whole
system, is zero. Depending on the values chosen for the parameters
of the Haldane model \textendash{} $M$ and $\phi$ \textendash{}
it is possible to obtain a trivial insulating phase ($C=0$) or topological
phases with $C=\pm1$, as can be seen in fig.~\ref{fig:phase_diagram}(a)
where we show the phase diagram of the Haldane model, reproduced here
for clarity and comparison. Believed to be of unlikely realization
in a solid state system, this model settled ground to the discovery
of time reversal invariant topological insulators \citep{KM05,bernevig2006quantum,BHZ2006,KWB+07},
which fuelled the field in the last decade \citep{Chiu2016}. The
observation of the quantum anomalous Hall effect was acheived in a
magnetic topological insulator \citep{qhaexp2013}. This was followed
by an experimental realization of the pioneer Haldane model using
cold atoms trapped in optical lattices \citep{Jotzu2014}, opening
the door to study the effect of interactions \citep{2018arXiv180800978G}
and disorder \citep{2018arXiv180711247G,PhysRevB.92.085410,PhysRevB.93.245414} in Chern insulators
experimentally in a controlled way.

In the presence of multiple energy bands and translational invariance,
a topological phase transition is associated to a gap closing at some
band crossing point(s) in the reciprocal space \citep{haldane2004berry,SF008}.
If the energy dispersion relation near these points is linear in the
reciprocal space vector $\bm{k}$, they are called Dirac points \citep{bernevigBook}.
Since Dirac points are monopoles of Berry curvature \citep{haldane2004berry},
they arise naturally in Chern insulators at the boundary between phases
with different Chern number. In the Haldane model, as recognized in
the seminal paper~\citep{Haldane1988}, there is a single Dirac point
along the full phase transition line shown in fig.~\ref{fig:phase_diagram}(a),
except at the point $M=0$, $\phi=0$, where there are two Dirac points
\textendash{} the case of graphene. Either one or two, Dirac points
in the Haldane model are bound to the corners of the hexagonal Brillouin
zone and they do not move by changing the parameters $M$ and $\phi$
of the model.

Under certain conditions, Dirac points can move, merge and split in
$\bm{k}$-space for continuous variations of the Hamiltonian parameters.
This can be achieved, for example, by considering models with asymmetric
changes in the hopping parameters \citep{Montambaux2009prb,Montambaux2009,Lim2012,Hasegawa2012,sticlet2013}.
In graphene, hopping changes may be induced by applying strain \citep{PhysRevB.80.045401,VKG10},
but the high strain values required strongly limit the effect. To
circumvent this limitation, moving and merging Dirac points have been
proposed in ac-driven graphene \citep{Delplace2013}, patterned graphene
\citep{Dvorak2015}, and artificial graphene \citep{Feilhauer2015}.
They have also been proposed in honeycomb optical lattices \citep{Wunsch2008,Sriluckshmy2014}
and square optical lattices \citep{Hou2014}, which seem to be even
better platforms. Indeed, the ability to create, move and merge Dirac
points \citep{Tarruell2012,Tarnowski2017}, has been recently demonstrated
using honeycomb optical lattices. Possible applications of merging
Dirac points range from valleytronics \citep{Ang2017} to plasmonics
\citep{Pyatkovskiy2016}. Interesting effects due to electron-electron
interactions \citep{Dora2013,Wang2013} and disorder have also been
proposed \citep{Carpentier2013}.

Here we propose new model for a Chern insulator realized on the square
lattice, with a richer phase diagram than the Haldane model, as shown
in fig.~\ref{fig:phase_diagram}(b). Two interesting properties stand
out: (i) a topological transition between $C=\pm1$ phases exists
for a finite staggered potencial range $-1<M<+1$, as depicted by
the vertical red line in fig.~\ref{fig:phase_diagram}(b), while
in the Haldane model such transition is reduced to the point $M=0$,
$\phi=0$; (ii) up to four Dirac points are found at the phase transition
lines, which are allowed to wander, merge, and split in reciprocal
space as a function of the model parameters $M$ and $\phi_{3}$ (to
be explained below), while in the Haldane model they are bound to
fixed momenta. The model can be simulated using a square optical lattice,
enabling the simultaneous analysis of Chern insulating phases and
their phase transitions, as well as the movement, merging and splitting
of Dirac points. Semi-Dirac points are also realized for some merging
conditions.

\section{Model and methods}

\subsection{A simple square lattice model}

\begin{figure}[H]
\begin{centering}
\includegraphics[width=0.8\columnwidth]{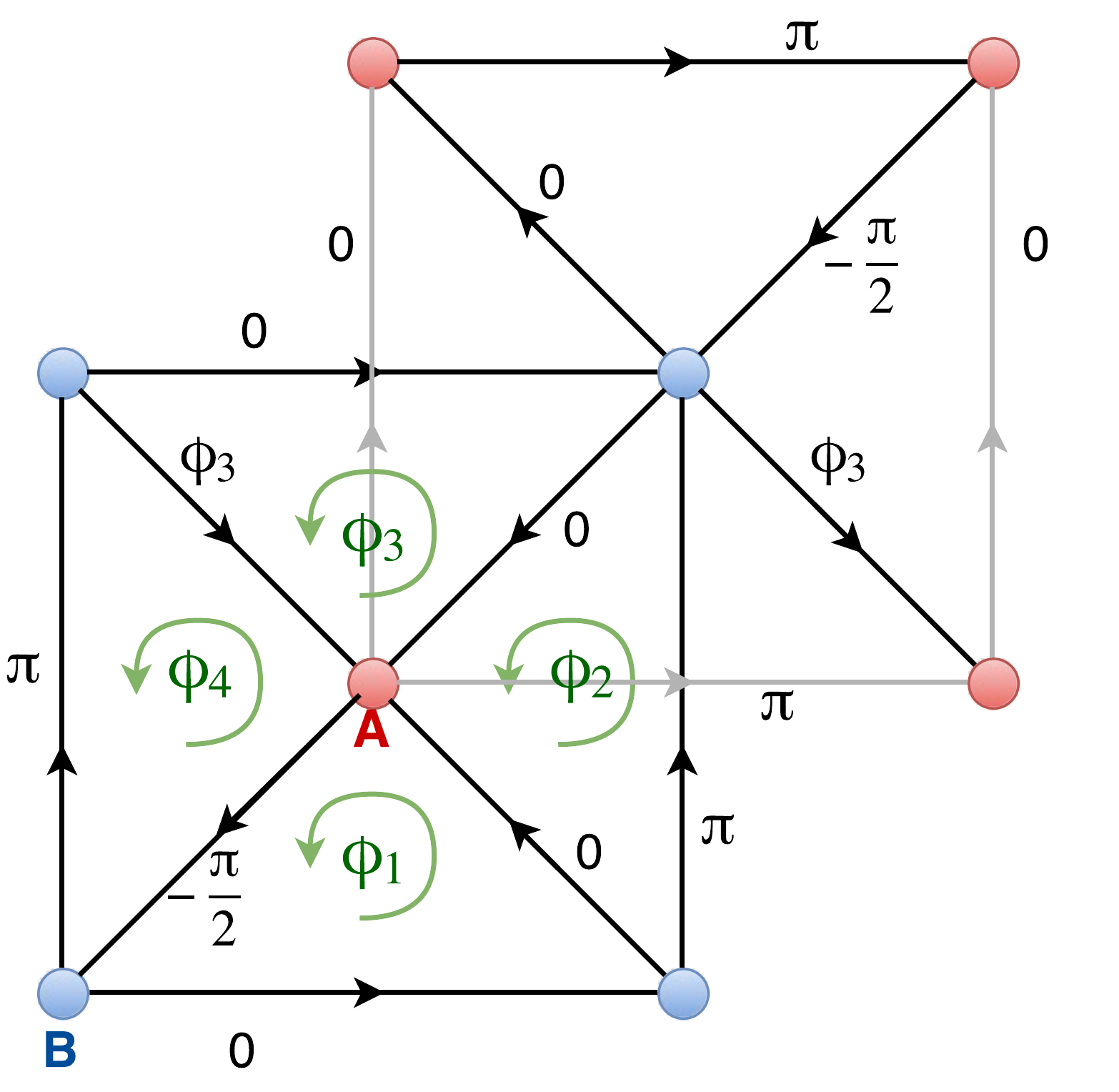}
\par\end{centering}
\caption{Representation of the lattice used for the toy model. The red and
blue atoms correspond respectively to different atoms of types A and
B. The complex hopping phases are given for each hopping and the flow
of the corresponding flux is represented by the arrows. \label{fig:lattice_model}}
\end{figure}

We present here a toy model composed by a lattice with two interpenetrating
square lattices of atoms~A and~B, as shown in fig.~\ref{fig:lattice_model}.
This model considers complex hoppings between first and second neighbors,
respectively of the type $t_{1}e^{i\varphi_{1}^{\alpha}}$ and $t_{2}e^{i\varphi_{2}^{\beta}}$,
with $\alpha$ and $\beta$ being respectively the four indexes of
the first and second neighbors. The phases $\varphi_{1}^{\alpha}$
and $\varphi_{2}^{\beta}$ were imposed in such a way that the total
flux in the unit cell, $\Phi_{T}$, is null. Designating $\phi_{i}$
as the flux over each of the four triangles composing the unit cell,
as represented in fig.~\ref{fig:lattice_model}, we impose $\Phi_{T}=\sum_{i=1}^{4}\phi_{i}=0$.
This condition does not univocally fix the phases, and we indicate
our choice for the phases $\varphi_{1}^{\alpha}$ and $\varphi_{2}^{\beta}$
explicitly in Fig.~\ref{fig:lattice_model}. 

The free parameters of our model were chosen to be the phase $\phi_{3}$
(and not $\phi$ to clearly distinguish from the fluxes in Haldane
model) represented in fig.~\ref{fig:lattice_model} and a sublattice
potential $M$: $+M$ for sites~A and $-M$ for sites~B. For the
present choice, the remaining fluxes are given by,
\begin{align}
\phi_{1}= & -\pi/2\nonumber \\
\phi_{2}= & \pi\nonumber \\
\phi_{4}= & -\pi/2-\phi_{3}\,.\label{eq:phi124}
\end{align}
The considered fluxes are asymmetric in that they explicitly break
the $C_{4}$ symmetry of the original lattice.

The Hamiltonian of the present model is easily written in direct space
using second quantization and the field operators $c_{\boldsymbol{r}}^{\dagger}$,
where $c_{\boldsymbol{r}}^{\dagger}=(a_{\boldsymbol{r}}^{\dagger},b_{\boldsymbol{r}}^{\dagger})$
is the creation operator in the sublattice pseudospin space ($a,b$
denoting the A and B sublattices, respectively) acting on the unit
cell at position $\boldsymbol{r}$. Taking into account translational
invariance and introducing the Bloch basis, $c_{\bm{r}}^{\dagger}=\frac{1}{\sqrt{N}}\sum_{\bm{{k}}}e^{-i\bm{{k.\bm{{r}}}}}c_{\bm{{k}}}^{\dagger}$,
we can write the Hamiltonian of our toy model in the reciprocal space
as $H=\sum_{\bm{k}}\Psi_{\bm{k}}^{\dagger}H(\bm{k})\Psi_{\bm{k}}$,
where $\Psi_{\bm{k}}^{\dagger}=(a_{\bm{k}}^{\dagger},b_{\bm{k}}^{\dagger})$
and

\begin{equation}
H(\bm{k})=\begin{bmatrix}g(\bm{k})+M & f(\bm{k})\\
f^{*}(\bm{k}) & -g(\bm{k})-M
\end{bmatrix}\,,\label{eq:Ham}
\end{equation}
with
\begin{align}
g(\bm{k})= & 2t_{2}[cos(\bm{k}.\bm{a_{1}})-cos(\bm{k}.\bm{a_{2}})]\,,\label{eq:gk}
\end{align}
\begin{align}
f(\bm{k})= & t_{1}[-i+e^{i\bm{k}.\bm{a_{1}}}+e^{-i\phi_{3}}e^{i\bm{k}.\bm{a_{2}}}+e^{i\bm{k}.(\bm{a_{1}}+\bm{a_{2}})}]\,,\label{eq:fk}
\end{align}
 and the primitive unit vectors $\bm{a_{1}}=a(1,0)$ and $\bm{a_{2}}=a(0,1)$,
written in terms of the lattice constant $a$.

It is useful to write the hamiltonian in terms of the Pauli matrices.
In units where the lattice constant $a=1$, it becomes

\begin{equation}
H(\bm{k})=\bm{h(\bm{k})}.\bm{\sigma}\,,\label{eq:hamgeneral}
\end{equation}
with the vector $\boldsymbol{h}$ given by
\begin{align}
h_{x}(\bm{k})= & t_{1}[\cos k_{x}+\cos(-\phi_{3}+k_{y})+\cos(k_{x}+k_{y})]\nonumber \\
h_{y}(\bm{k})= & t_{1}[1-\sin k_{x}-\sin(-\phi_{3}+k_{y})-\sin(k_{x}+k_{y})]\nonumber \\
h_{z}(\bm{k})= & 2t_{2}(\cos k_{x}-\cos k_{y})+M\,,\label{eq:hvec}
\end{align}
and $\boldsymbol{\sigma}=(\sigma_{x},\sigma_{y},\sigma_{z})$ is the
vector of Pauli matrices. From eq.~\eqref{eq:hamgeneral}, the band
spectrum is simply obtained as $\epsilon(\bm{k})=\pm|\bm{h}|$. As
shown below, analytic expression for band crossing points can be obtained
by solving $\epsilon(\bm{k})=0$. 

When the Hamiltonian is in the form of eq.~\eqref{eq:hamgeneral},
the Chern number $C$ can be calculated using the following expression
\citep{Xiao2010},

\begin{equation}
C=\frac{1}{4\pi}\int_{\mathrm{BZ}}\frac{\partial\bm{h}}{\partial k_{x}}\times\frac{\partial\bm{h}}{\partial k_{y}}\cdot\frac{\bm{h}}{|h|^{3}}\,dk_{x}dk_{y}\,,\label{eq:chern_number}
\end{equation}
where the integration is performed over the Brillouin zone (BZ): $k_{x},k_{y}\in[-\pi,\pi]$.

\subsection{Low energy, continuum approximation}

\label{subsec:low_energy_intro}

In the presence of a topological phase transition, the spectral gap
existing in the insulating phase closes at some band crossing points.
These gap closing points are responsible for the change in the Chern
number between the phases. Near these points the band crossing is
generally linear \citep{haldane2004berry}, and we are able to linearize
the Hamiltonian in eq.~\eqref{eq:hamgeneral} provided we are close
enough to the phase transition.

Let us assume that a band crossing point exists at point P in the
BZ associated to a given phase transition in the phase diagram. After
linearization around P, assuming the parameters are such that the
system is close to the phase transition, we may approximate eq.~\eqref{eq:hamgeneral}
by 

\begin{equation}
H^{\mathrm{P}}(\bm{\kappa})=\kappa_{\alpha}\mathcal{G}_{\alpha\beta}^{\mathrm{P}}\sigma_{\beta}\,,\label{eq:Gab}
\end{equation}
where $\bm{\kappa}\equiv(\kappa_{x},\kappa_{y})=\bm{k}-\bm{k}_{\mathrm{P}}$
is the small momentum relative to $\boldsymbol{k}_{\mathrm{P}}$,
with $\boldsymbol{k}_{\mathrm{P}}$ the momentum associated to the
band crossing point P. In eq.~\eqref{eq:Gab}, summation over repeated
indeces is assumed, with $\kappa_{z}=1$, and the matrix $\mathcal{G}^{\mathrm{P}}$
is the matrix obtained by expanding the Hamiltonian in eq.~\eqref{eq:hamgeneral}
around a given band crossing point P. Due to the linear form of the
approximate Hamiltonian in eq.~\eqref{eq:Gab}, these gap closing
points are called \emph{Dirac points}.

The contribution of the Dirac point P to the Chern number, $C_{\mathrm{P}}$,
is obtained using eq.~\eqref{eq:chern_number} and~\eqref{eq:Gab},

\begin{equation}
C_{\mathrm{P}}=\frac{1}{2}\text{sgn}[\det(\mathcal{G}^{\mathrm{P}})]\,.\label{eq:chern_point}
\end{equation}
If we obtain $C_{\mathrm{P}}$ in two distinct phases close to the
phase transition, we may then compute the difference between the two
results, $\Delta C_{\mathrm{P}}$. Since the high energy contributions
cancel, we obtain the total contribution of this Dirac point to the
change in Chern number at the phase transition \citep{Lovesey2014}.
We use $\Delta C_{\mathrm{P}}$ to further characterize the phase
diagram of the proposed Chern insulator.

\section{Results}

\subsection{Phase diagram}

\label{subsec:phase_diagram}

The obtained phase diagram is shown in fig.~\ref{fig:phase_diagram}(b).
It was obtained through eq.~\eqref{eq:chern_number} and confirmed
numerically with the Fukui method \citep{Fukui2005}.

The $C=0$ phase corresponds to a normal insulator. The energy gap
closes at the dashed and dotted curves which correspond to phase transition
lines between a trivial and a topological phase. There are two topological
phases with $C=\pm1$ corresponding to the filled regions in fig.~\ref{fig:phase_diagram}(b).
They are separated by a vertical phase transition line existing at
$\phi_{3}=\pi/2$. The equations for the dotted, dashed, and solid
phase transition lines shown in fig.~\ref{fig:phase_diagram}(b)
are

\begin{equation}
\begin{cases}
M(\phi_{3})=\pm2\sqrt{2}t_{2}\cos(\phi_{3}/2) & \text{{dotted(+)/dashed(-)}}\\
\phi_{3}=\pi/2\cap\Big|\frac{M}{2t_{2}}\Big|<1 & \text{{solid\,(vertical)}}\,.
\end{cases}
\end{equation}

There is yet another line not depicted in fig.~\ref{fig:phase_diagram}(b),
correponding to $\phi=\pi/2$ and $-4t_{2}<M<-2t_{2}$, which does
not correspond to any phase transition but for which the energy spectrum
is gapless. 

\begin{figure*}
\begin{centering}
\includegraphics[width=0.95\textwidth]{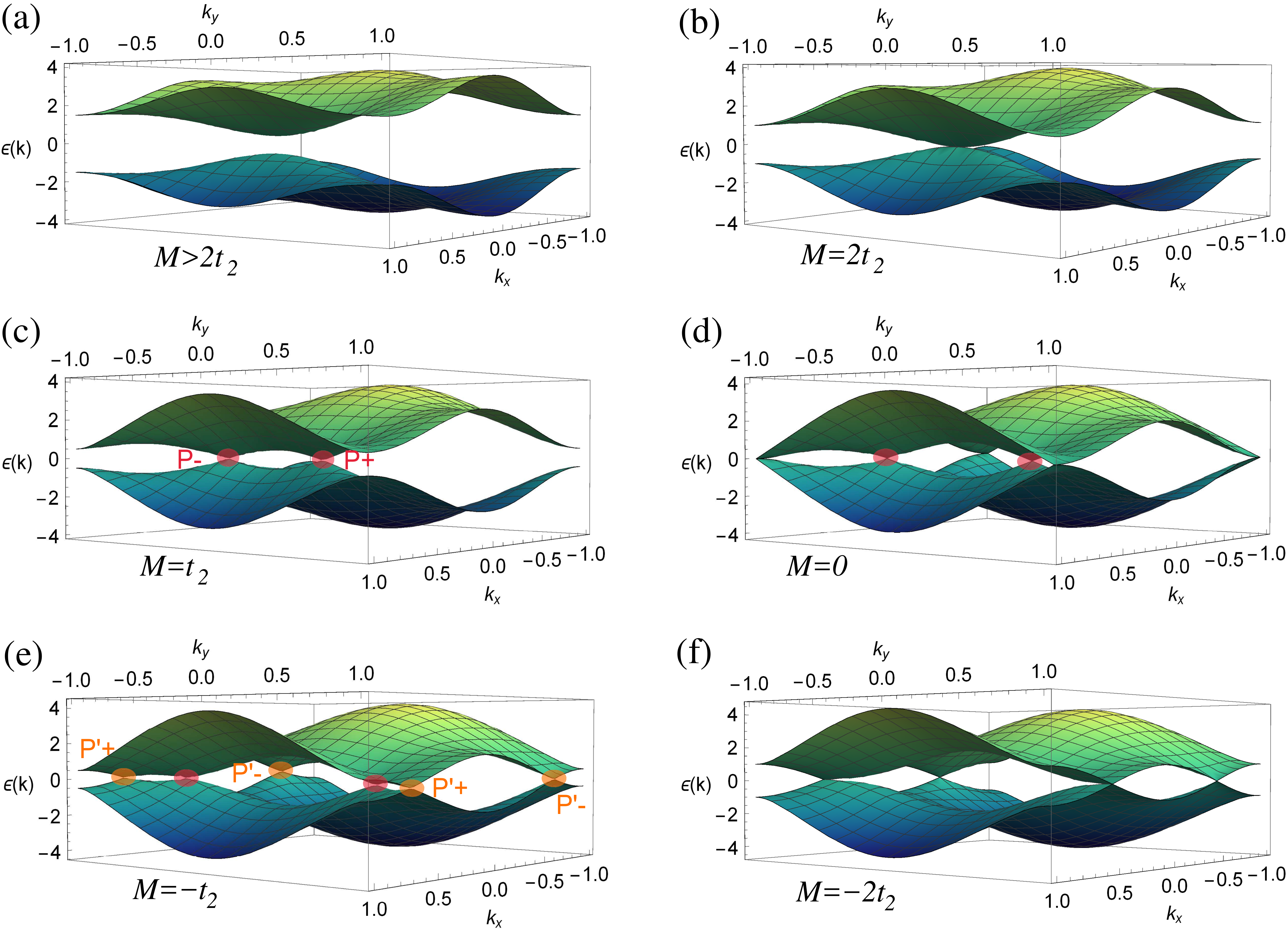}
\par\end{centering}
\caption{Motion of the Dirac points along the $\phi_{3}=\pi/2$ phase transition
line: (a) $M>2t_{2}$, (b) $M=2t_{2}$, (c) $M=t_{2}$, (d) $M=0$,
(e) $M=-t_{2}$, (f) $M=-2t_{2}$.\label{fig:dirac_points_motion_spectrum}}
\end{figure*}

\subsection{Wandering of Dirac points and its consequences}

\label{subsec:wandering_dirac_points}

We have analyzed the evolution of the number of Dirac points, as well
as their position in the BZ, along the phase transition lines shown
in fig.~\ref{fig:phase_diagram}(b). We have verified that the position
of the Dirac points changes along the transition lines. Furthermore,
the number of band crossing points is not conserved. In this section
we provide a detailed analysis of this behavior.

\subsubsection{Single band crossing point}

The dotted curve in fig.~\ref{fig:phase_diagram}(b) is associated
with a single band crossing point everywhere, except at $\phi_{3}=\pm\pi$
where there are two Dirac points. The same applies to the dashed curve,
except at $\phi_{3}=\pi/2$, where it also has two band crossing points.
For for $\phi_{3}\neq\{\pm\pi,\pi/2\}$, the position of the single
Dirac point along these curves is given, for the dotted curve, by

\begin{equation}
\begin{cases}
(k_{x},k_{y})=(x_{M_{+}},y_{M_{+}}) & \phi_{3}<0\\
(k_{x},k_{y})=(x_{M_{-}},y_{M_{-}}) & \phi_{3}>0
\end{cases}\,,
\end{equation}
and for the dashed curve, by

\begin{equation}
\begin{cases}
(k_{x},k_{y})=(x_{M_{-}},y_{M_{-}}) & \phi_{3}<0\\
(k_{x},k_{y})=(x_{M_{+}},y_{M_{+}}) & \phi_{3}>0
\end{cases}\,,
\end{equation}
where

\begin{equation}
\begin{cases}
x_{M_{\pm}}=\pm2\text{arctan}\left[\frac{M+2t_{2}}{\sqrt{8t_{2}^{2}-M^{2}}\pm2t_{2}}\right]\\
y_{M_{\pm}}=2\text{arctan}\left[\frac{\sqrt{8t_{2}^{2}-M^{2}}\mp M}{\sqrt{8t_{2}^{2}-M^{2}}\pm(M+4t_{2})}\right]
\end{cases}\,.\label{eq:xMyM}
\end{equation}

\subsubsection{Multiple band crossing points}

At $\phi_{3}=\pm\pi$, we have two Dirac points at $(k_{x},k_{y})=(-3\pi/4,-3\pi/4)$
and $(k_{x},k_{y})=(\pi/4,\pi/4)$. For $\phi_{3}=\pi/2$, the phase
transition line is very rich in terms of band crossing points as their
number ranges between 1 and 4 due to merging and splitting of Dirac
points. This case is analyzed in the following.

Let us fix $\phi_{3}=\pi/2$. As long as $M>2t_{2}$, we are not yet
at the phase transition line and there is a gap in the energy spectrum
corresponding to a trivial insulator {[}fig.~\ref{fig:dirac_points_motion_spectrum}(a){]}.
This gap closes for $M=2t_{2}$, at a single band crossing point,
$(k_{x},k_{y})=(\pi/2,0)$, {[}fig.~\ref{fig:dirac_points_motion_spectrum}(b){]}.
This band crossing point is not a conventional Dirac point, as the
energy spectrum is quadratic in the $k_{y}$ direction and linear
in $k_{x}$ \textendash{} which is called a semi-Dirac point \citep{Montambaux2018}.
This is known to be a consequence of the merging of two Dirac points
\citep{Montambaux2009,Montambaux2009prb,pardoSemDirac,Li2015,Adroguer2016},
in this case in the $k_{y}$ direction. Indeed, topological phase
transitions accompained with higher order band crossings, for example
quadratic band crossings, exist and can be interpreted as the merging
of two or more Dirac points \citep{SYF+09}.

Slightly reducing $M$, still with fixed $\phi_{3}=\pi/2$, we observe
that the semi-Dirac point splits in the above mentioned two Dirac
points in the $k_{y}$ direction {[}fig.~\ref{fig:dirac_points_motion_spectrum}(c){]},
which we call $\mathrm{P}_{+}$ and $\mathrm{P}_{-}$. These points
move away from each other in the $k_{y}$ direction, untill they reach
the boundaries of the BZ for $M=-2t_{2}$ {[}fig.~\ref{fig:dirac_points_motion_spectrum}(f){]},
where they again merge. Their coordinates for $-2t_{2}<M<2t_{2}$
are given by
\begin{equation}
P_{\pm}=\left\{ \frac{\pi}{2},\pm\text{\ensuremath{\arccos}}\left(\frac{M}{2t_{2}}\right)\right\} \,.\label{eq:pp}
\end{equation}
For $M=0$ {[}fig.~\ref{fig:dirac_points_motion_spectrum}(d){]},
a new band crossing point shows up at the boundaries of the BZ, $(k_{x},k_{y})=(\pm\pi,\pm\pi)$.
This point is also a semi-Dirac point, being quadratic in the $k_{x}$
direction. For $-2t_{2}<M<0$ {[}fig.~\ref{fig:dirac_points_motion_spectrum}(e){]},
this single point splits in two in the $k_{x}$ direction, which we
call $P'_{+}$ and $P'_{-}$, with coordinates given by
\begin{equation}
P'_{\pm}=\left\{ \pi,\pm\text{arccos}\left(-\frac{M}{2t_{2}}-1\right)\right\} \,.\label{eq:pprime}
\end{equation}
An important detail not shown in the figures is that these points
exist for $-4t_{2}<M<0$, and therefore survive outside the boundaries
of the $\phi_{3}=\pi/2$ phase transition line. They give rise to
the gapless spectrum associated with the extension of the vertical
transition line in fig.~\ref{fig:phase_diagram}(b), mentioned in
section~\ref{subsec:phase_diagram}.

\begin{figure}
\centering{}\includegraphics[scale=0.7]{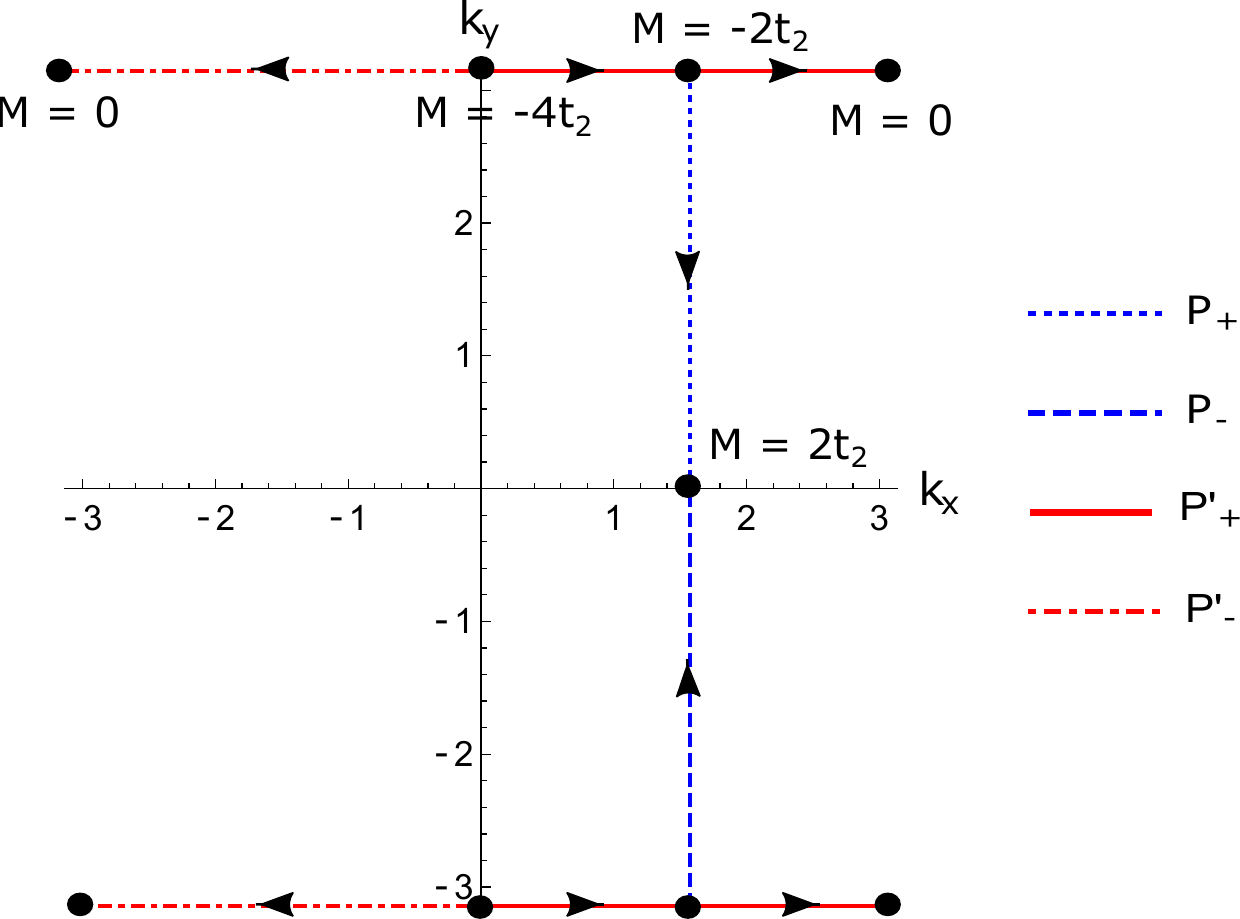}\caption{Overview of the motion, merging and splitting of the Dirac points
in the BZ for $\phi_{3}=\pi/2$ and $-4t_{2}<M<2t_{2}$ {[}along the
vertical line in Fig.~\ref{fig:phase_diagram}(b){]}. All four Dirac
points existing in this region, $P_{\pm}$ and $P'_{\pm}$ {[}eq.~\eqref{eq:pp}
and~\eqref{eq:pprime}{]}, are depicted in the figure with the corresponding
motion lines given in the legend. The arrows point to regions of higher
$M$ and the dots describe the merging or splitting of Dirac points.\label{fig:dirac_point_motion_overview}}
\end{figure}

A global overview of the motion, merging and splitting of the Dirac
points for the $\phi=\pi/2$ gapless region of the phase diagram ($-4t_{2}<M<2t_{2}$)
is given in figure \ref{fig:dirac_point_motion_overview}. 

\subsubsection{Contribution to the topological invariant}

Based on the methods from section~\ref{subsec:low_energy_intro},
we computed the contributions of Dirac points to the Chern number
for the different phase transition curves. For curves associated to
a single Dirac point {[}dashed and dotted in fig.~\ref{fig:phase_diagram}(b){]},
its contribution was confirmed to be trivially $\Delta C=1$. 

For the $\phi_{3}=\pi/2$ phase transition line the contributions
are as follows. By expanding the Hamiltonian in eq.~\eqref{eq:hamgeneral}
around $P_{+}$, the contribution to the Chern number obtained from
eq.~\eqref{eq:chern_point} is

\begin{equation}
C_{P_{+}}=\begin{cases}
-1/2 & \phi_{3}\lesssim\pi/2\\
1/2 & \phi_{3}\gtrsim\pi/2
\end{cases}\,.
\end{equation}
This meas that $P_{+}$ contributes with $\Delta C_{P_{+}}=1$ for
the change in the Chern number. Doing the same for $P_{-}$, we verified
that this point had exactly the same contribution. Therefore, $P_{+}$
and $P_{-}$ contribute to a total change of $\Delta C=2$ between
the topological phases. For the points $P'_{+}$ and $P'_{-}$, we
obtained that for all their domain of existence

\begin{equation}
C_{P'_{+}}=-C_{P'_{-}}=\begin{cases}
1/2 & \phi_{3}\lesssim\pi/2\\
-1/2 & \phi_{3}\gtrsim\pi/2
\end{cases}\,,
\end{equation}
thus giving canceling contributions. We then conclude that $P_{+}$
and $P_{-}$ are the points responsible for the change from $C=-1$
to $C=1$ between topological phases. It thus makes sense that $P_{+}$
and $P_{-}$ exist for $-2t_{2}<M<2t_{2}$, while and the additional
points $P'_{+}$ and $P'_{-}$ exist for $-4t_{2}<M<0$. If the latter
were to have any net contribution for the topological transition,
we would have a phase transition for $-4t_{2}<M<2t_{2}$ which is
not the case. 

\section{Conclusions}

We have put forward a new model for a Chern insulator which is characterized
by a richer phase diagram than the celebrated Haldane Chern insulator.
An interesting new result regarding the obtained phase diagram is
the appearance of a phase transition line between topological phases
in the $(M,\phi_{3})$ parameter space. This implies a transition
between non-trivial phases for non-zero sublattice potential. This
phenomenon does not occur for the Haldane model in which the transition
between $C=\pm1$ phases occurs only for $M=0$ {[}compare fig.~\ref{fig:phase_diagram}(a)
with~\ref{fig:phase_diagram}(b){]}. 

Another striking characteristic of the proposed Chern insulator is
that Dirac points at the phase transition lines can move, merge, and
split in reciprocal space, while in the Haldane model they are bound
to fixed momenta. Of particular relevance is the behavior of Dirac
points for the vertical transition line in fig.~\ref{fig:phase_diagram}(b),
when $\phi_{3}=\pi/2$. There, up to four Dirac points are found,
which move as $M$ is changed, merging in pairs as semi-Dirac points,
with subsequent annihilation or splitting. Of the four Dirac points
found, only two contribute to the change $\Delta C=2$ in Chern number
across the phase transition, where $C=+1$ on one side of the transition
line and $C=-1$ on the other. The phase transition exists as long
as both points are separated and ceases to exist when they merge and
a trivial insulating gap is opened.

The model here proposed extends the physics of moving and merging
Dirac points, studied previously in models with tunable hopping values
\citep{Wunsch2008,Montambaux2009,Hou2014}, to the realm of Chern
insulators where only hopping phases ($\phi_{3}$ in the present case),
and not their absolute values, have to be changed. It also enables
the study of semi-Dirac points \citep{Li2015} within the same set
up where Chern insulating phases are realized.

The realization of the Haldane Chern insulator \citep{Jotzu2014},
as well as the ability to create, move and merge Dirac points \citep{Tarruell2012,Tarnowski2017},
has been recently demonstrated using ultracold atoms in an optical
lattice. In Refs.~\citep{Jotzu2014} and~\citep{Tarruell2012, Tarnowski2017},
a challenging honeycomb lattice to trap fermionic atoms had to be
realized using interfering laser beams. Within the present model,
a much simpler square lattice is required, thus facilitating the study
of Chen insulating properties and Dirac points merging and wandering,
which can be done simultaneously within the same setting. It would
also be interesting to study the effect of interactions in such a
rich playground \citep{Cocks2012}.
\begin{acknowledgments}
The authors acknowledge partial support from FCT-Portugal through
Grant No. UID/CTM/04540/2013. PR acknowledges support by FCT-Portugal
through the Investigador FCT contract IF/00347/2014. 
\end{acknowledgments}

\bibliographystyle{apsrev}
\bibliography{mybib}

\end{document}